\documentclass[12pt]{article}
\usepackage[mathscr]{eucal}
\usepackage{amssymb,latexsym}
\usepackage{verbatim}
\usepackage{graphicx}
\usepackage{citesort}
\usepackage{amsmath}
\usepackage{amsthm}
\usepackage{enumerate}
\usepackage{authblk}
\usepackage{color}
\usepackage{url}

\usepackage{caption}

\usepackage[normalem]{ulem}

\newtheorem{thm}{Theorem}[section]
\newtheorem{lem}[thm]{Lemma}

\renewcommand\l{\lambda}

\newcommand\bbR{{\mathbb R}}

\renewcommand\S{\Sigma}

\newcommand\s{\sigma}

\newcommand\e{\epsilon}

\renewcommand\div{{\rm div}}

\renewcommand\l{\lambda}
\newcommand\g{\gamma}

\renewcommand\th{\theta}

\newcommand\beq{\begin{equation}}
\newcommand\eeq{\end{equation}}
\newcommand\ben{\begin{enumerate}}
\newcommand\een{\end{enumerate}}
\newcommand\bit{\begin{itemize}}
\newcommand\eit{\end{itemize}}

\DeclareMathOperator{\diver}{div}

\renewcommand{\div}{\diver}

\newcommand{\R}{\mathbb R}

\newcommand{\ov}{\overline}

\newcommand{\pd}{\partial}

\DeclareFontFamily{OT1}{rsfs}{} \DeclareFontShape{OT1}{rsfs}{m}{n}{ <-7> rsfs5 <7-10> rsfs7 <10-> rsfs10}{}
\DeclareMathAlphabet{\mycal}{OT1}{rsfs}{m}{n}
\def\scri{{\mycal I}}
\newcommand{\mcD}{{\mycal D}}

\newcommand{\mcM}{{\mycal M}}
\newcommand{\bmcD}{\,\,\widetilde{\!\!\mcD}}

\newcommand{\bmcM}{\,\,\,\,\widetilde{\!\!\!\!\mcM}}

\newcounter{mnotecount}

\title{Weakly trapped surfaces in \\ asymptotically de Sitter spacetimes}

\author[1,3]{Piotr T. Chru\'sciel}
\author[2,3]{Gregory J. Galloway}
\author[2]{Eric Ling}

\affil[1]{Gravitational Physics,
University of Vienna,  Austria}

\affil[2]{Department of Mathematics

University of Miami, Coral Gables, FL, USA}

\affil[3]{Erwin Schr\"odinger Institute,
University of Vienna,  Austria}

\begin{document}
\date{}
\maketitle
\vspace{.2in}

\begin{abstract}
It is a standard fact that trapped or marginally trapped surfaces are not visible from conformal infinity, under the usual set of conditions on matter fields and the conformal completion, provided that the cosmological constant is non-positive. In this note we show that the situation is more delicate in the presence of a positive cosmological constant: we present examples of visible marginally trapped surfaces, and we provide a set of natural conditions which guarantee non-visibility.
\end{abstract}



\section{Introduction}

A classical result in the theory of black holes asserts that trapped surfaces are, in a suitable sense, `externally invisible'. Somewhat more precisely,  for spacetimes $(\mcM,g)$ which are asymptotically flat (in the sense of admitting a suitably regular future null infinity $\scri^+$) and which satisfy appropriate energy and causality conditions, no (future)  trapped surface ($\th^{\pm} < 0$) can be contained in $I^-(\scri,^+ \bmcM)$, where
$\bmcM = \mcM \cup \scri^+$.    In fact, this result also extends to (future) weakly trapped  surfaces ($\th^{\pm} \le 0$). Indeed, using a suitable notion of global hyperbolicity, the following has been shown in~\cite{CGS}:

\begin{thm}[\cite{CGS}, Theorem 6.1]\label{notrapped0}
Let $(\mcM,g)$ be an
 asymptotically flat spacetime, in the sense of admitting a
 regular future conformal completion $\bmcM = \mcM \cup \scri^+$,
 where $\scri^+$ is a connected null hypersurface, such that,
 \begin{enumerate}
 \item $\bmcD = \mcD \cup \scri^+$ is globally hyperbolic, where $\mcD = I^-(\scri^+,\bmcM)$, and
 \item  for any compact set $K \subset \mcD$, $J^+(K, \bmcD)$ does not
 contain all of~$\scri^+$ (``$i^0$-avoidance").
  \end{enumerate}
 If the  null energy condition holds on $\mcD$  then there are no  weakly trapped surfaces
 within $\mcD$.
\end{thm}

\smallskip

Since   spacetimes with a positive cosmological constant are of considerable current interest (cf., e.g., \cite{StromingerdS,Riess:2004nr,AshtekarBK}), it is useful to have an analogue of Theorem~\ref{notrapped0} for asymptotically de Sitter spacetimes. We provide such a result in Theorem~\ref{notrapped in dS} below.

To put our result in proper context, it should be kept in mind that weakly trapped surfaces occur in the past of $\scri^+$ in asymptotically de Sitter spacetimes.  We review some well known and not so well known examples in Section~\ref{Examples}.  One sees from these examples that the visibility of future trapped surfaces in this case is due to the failure of a suitable analogue of the $i^0$-avoidance condition in Theorem \ref{notrapped0}.  When this is taken into account, an analogue of Theorem~\ref{notrapped0} can be obtained, and under a causality condition imposed only on the physical spacetime $\mcM$.

Indeed, Theorem \ref{main} and Lemma \ref{lemma} below immediately yield our main result (see below for definitions and terminology):

\smallskip

\begin{thm}\label{notrapped in dS}
Consider a future asymptotically de Sitter spacetime  $(\mcM,g)$ which is future causally simple and  satisfies  the null energy condition.  Let $A \subset \mcM$ be such that $J^+(A, \bmcM)$ does not contain all of $\scri^+$.   Then there are no future weakly trapped surfaces contained in $J^+(A, \bmcM) \cap I^-(\scri^+, \bmcM)$.\
\end{thm}

\section{Invisibility of trapped surfaces in the asymptotically de Sitter setting}
\label{results}

Let $(\mcM,g)$ be a spacetime and consider a compact co-dimension two spacelike submanifold $S \subset \mcM$. Under appropriate orientability assumptions, there exists a smooth unit spacelike vector field $\nu$ normal to $S$.
From this one obtains unique future directed null vector fields $l^\pm$ normal to $S$ such that $g(\nu,l^\pm) = \pm1$. These induce the null second fundamental forms $\chi^{\pm}\colon T_pS \times T_pS \to \R$ given by $\chi^{\pm}(X,Y) = g(\nabla_X l^{\pm}, Y)$. Tracing $\chi_{\pm}$, we obtain the null expansion scalars (or null mean curvatures) $\theta^{\pm} = {\rm tr}_S\chi^{\pm} = \div_Sl^{\pm}$. $S$ is \emph{future trapped} if both $\theta^{\pm} < 0$ and \emph{weakly future trapped} if both $\theta^{\pm} \leq 0$. If $\theta^{\pm} = 0$, then $S$ is \emph{marginally future trapped}.

A spacetime $(\mcM,g)$ will be said to admit  a \emph{future conformal completion} $(\bmcM, \tilde{g})$ provided $(\bmcM, \tilde{g})$ is a spacetime with boundary such that the following properties hold
\begin{enumerate}

\item[(a)] $\mcM$ is the interior of $\bmcM$,

\item[(b)] $\scri^+ := \pd \bmcM$ is smooth, connected, and lies to the future of $\mcM$, i.e. $\scri^+ \subset I^+(\mcM, \bmcM)$,

\item[(c)] There is a smooth function $\Omega$ on $\bmcM$ such that on $\mcM$ the metric $\tilde{g}$ is related to $g$ via $\tilde{g} = \Omega^2g$ where $\Omega = 0$ on $\scri^+$ and $d\Omega$ is nowhere vanishing on $\scri^+$.

\end{enumerate}
If in addition to these we also have that $\scri^+$ is spacelike, then we say $(\mcM, g)$ is \emph{future asymptotically de Sitter}. (We will sometimes omit the adjective ``future'' 
for conciseness.) In this case (b) above implies that $\scri^+$ is acausal. We say that $(\mcM,g)$ is \emph{future causally simple} provided $J^+(K,\mcM)$ is closed for all compact $K \subset \mcM$. Note that we do not assume any further causality conditions on $(\mcM, g)$. This agrees with \cite{HE} but differs from \cite{BEE} and \cite{MiSa}. In addition we say that the conformal completion $(\bmcM, \tilde{g})$ is \emph{future causally simple with respect to $\mcM$} provided $J^+(K, \bmcM)$ is closed for all compact $K \subset \mcM$.
As discussed in \cite{GalSol} (cf. Lemma 2.3), a spacetime with boundary $(\bmcM,\tilde{g})$ admits an extension to a spacetime $(\mcM',g')$ without boundary such that $\scri^+$ separates $\mcM'$. The existence of such an extension is assumed in what follows.

We first prove an invisibility theorem which uniformly covers the cases when $\scri^+$ is either a timelike, null, or spacelike hypersurface. In this theorem we assume $(\bmcM, \tilde{g})$ is future causally simple with respect to $\mcM$.  (The main use of the assumption of global hyperbolicity in the proof of Theorem 6.1 in \cite{CGS} was to ensure this.)

\medskip

\begin{thm}\label{main}
Suppose $(\mcM,g)$ satisfies the null energy condition and admits a future conformal completion $(\bmcM, \tilde{g})$ which is future causally simple with respect to $\mcM$. Suppose $\scri^+$ is either timelike, null, or spacelike. Let $A \subset \mcM$ be such that
$J^+(A, \bmcM)$ does not contain all of $\scri^+$. Then there are no future weakly trapped surfaces in $J^+(A, \bmcM) \cap I^-(\scri^+, \bmcM)$.

\end{thm}

\proof
Seeking a contradiction, suppose there exists a future weakly trapped surface $S \subset J^+(A, \bmcM) \cap I^-(\scri^+, \bmcM)$ for some set $A \subset \mcM$ such that $J^+(A, \bmcM)$ does not contain all of $\scri^+$. In particular this means $J^+(S,\bmcM)$ does not contain all of $\scri^+$. Therefore there exists a point
\[
q_0 \in \pd\big(J^+(S, \bmcM) \cap \scri^+ \big) =  \pd J^+(S, \bmcM) \cap \scri^+  \,,
\]
Introduce a Riemannian metric $h$ on $\scri^+$, and let $U$ be a convex normal neighborhood about $q_0$ in $\scri^+$. Let $q_1$ be a point in $U \setminus \pd J^+(S,\bmcM)$ (which is open in $U$), chosen so that some points of $J^+(S,\bmcM)$ lie within the injectivity radius of $q_1$.
Let $q \in {U} \cap \pd J^+(S,\bmcM)$ minimize the $h$-distance between $\pd J^+(S,\bmcM)$ and $q_1$ in $\ov{U}$. Let $r > 0$ be this minimizing distance.  Let $S^+$ be the geodesic ball with radius $r$ centered at $q_1$. Then $S^+$ is a smooth hypersurface in $\scri^+$ which contains $q$ and does not meet $I^+(S, \bmcM)$.

Since $(\bmcM,g)$ is future causally simple with respect to $\mcM$, the boundary of $J^+(S, \bmcM)$ satisfies
\[
\pd J^+(S, \bmcM) = J^+(S, \bmcM) \setminus I^+(S, \bmcM).
\]
Thus there exists a null geodesic $\g\colon [a,b] \to \bmcM$ such that $\g(a) \in S$, $\g(b) = q$, and emanating orthogonally from $S$. Since $\scri^+$ is either timelike, null, or spacelike, $\gamma$ must intersect $\scri^+$ transversally,\footnote{Without some control on the causal character of $\scri^+$,
this in general may fail.}  and hence, $\g\big([a,b)\big) \subset \mcM$. Moreover, since $\g$ does not enter the timelike future of $S$, there are no null focal points to $S$ along $\g$.
Then \cite[Theorem B.7]{SolThesis} implies that for all points along $\g$
the null exponential map has full rank. This allows one to generate a smooth null hypersurface $N_1 \subset J^+(S,\bmcM)$ containing the segment of $\g\big|_{[a,b-\e]}$ for $\e>0$ arbitrarily small. Since $S$ is future weakly trapped, the null future inwards and outwards mean curvatures $\theta^{\pm}$ of $S$ are nonpositive. Let $\theta_1 = \theta_1(s)$ be the null mean curvature of $N_1$ along $\g$. Hence $\theta_1(a) \leq 0$, then by the Raychaudhuri equation and the null energy condition, $\theta_1(s) \leq 0$ for all $s \in [a, b-\e]$.

There exists a future directed outward pointing null vector $\tilde{K}$ at $q$ orthogonal to $S^+$ in the unphysical metric $\tilde{g} = \Omega^2 g$. Since $\Omega \to 0$ as one approaches $\scri^+$, we see that - irregardless of the causal character of $\tilde{\nabla}\Omega$ at $q$ - we can normalize $\tilde{K}$ so that along $\gamma$ near $q$, we have

\[
\tilde{K}(\Omega) = \tilde{g}(\tilde{K}, \tilde{\nabla}\Omega) = -1  \,.
\]
Since $S^+$ does not meet $I^+(S,\bmcM)
$, we see that $\g$ intersects $S^+$ orthogonally. Let $N_2$ be a smooth null hypersurface which is a subset of $\pd J^-(S^+,\bmcM)$ and contains $\g\big([a,b)\big)$ near $q$. $\tilde{K}$ extends to a smooth null vector field on $S_2$ near $q$. The null mean curvature $\theta_2$ of $N_2$ with respect to the physical metric $g$ is related to the null mean curvature $\tilde{\theta}_2$ in the unphysical metric $\tilde{g}$ by
\begin{equation}\label{divergence relationships}
\theta_2 = -(n-1) \tilde{K}(\Omega) + \Omega \tilde{\theta}_2
\end{equation}
Our construction of $N_2$ from $S^+$ shows that $\tilde{\theta}_2$ is bounded on $N_2$. Therefore close to $q$ on $N_2$, we have $\theta_2 > 0$. However, since $S^+$ does not meet $I^+(S, \bmcM)$, we know that $N_1$ lies to the future side of $N_2$ on points of $\g$ near $q$. This contradicts the maximum principle for smooth null hypersurfaces.
\qed

\medskip

In the asymptotically de Sitter setting, the causal simplicity of $(\mcM,g)$ implies $(\bmcM,\tilde{g})$ is causally simple with respect to $\mcM$.

\medskip

\begin{lem}\label{lemma}
Suppose $(\mcM,g)$ is future asymptotically de Sitter with conformal completion $(\bmcM, \tilde{g})$. If $(\mcM,g)$ is future causally simple, then $(\bmcM,\tilde{g})$ is future causally simple with respect to $\mcM$.
\end{lem}

\smallskip

\newtheorem{example}[thm]{Example}

\begin{figure}
\begin{center}
\mbox{
\includegraphics[width=3.7in]{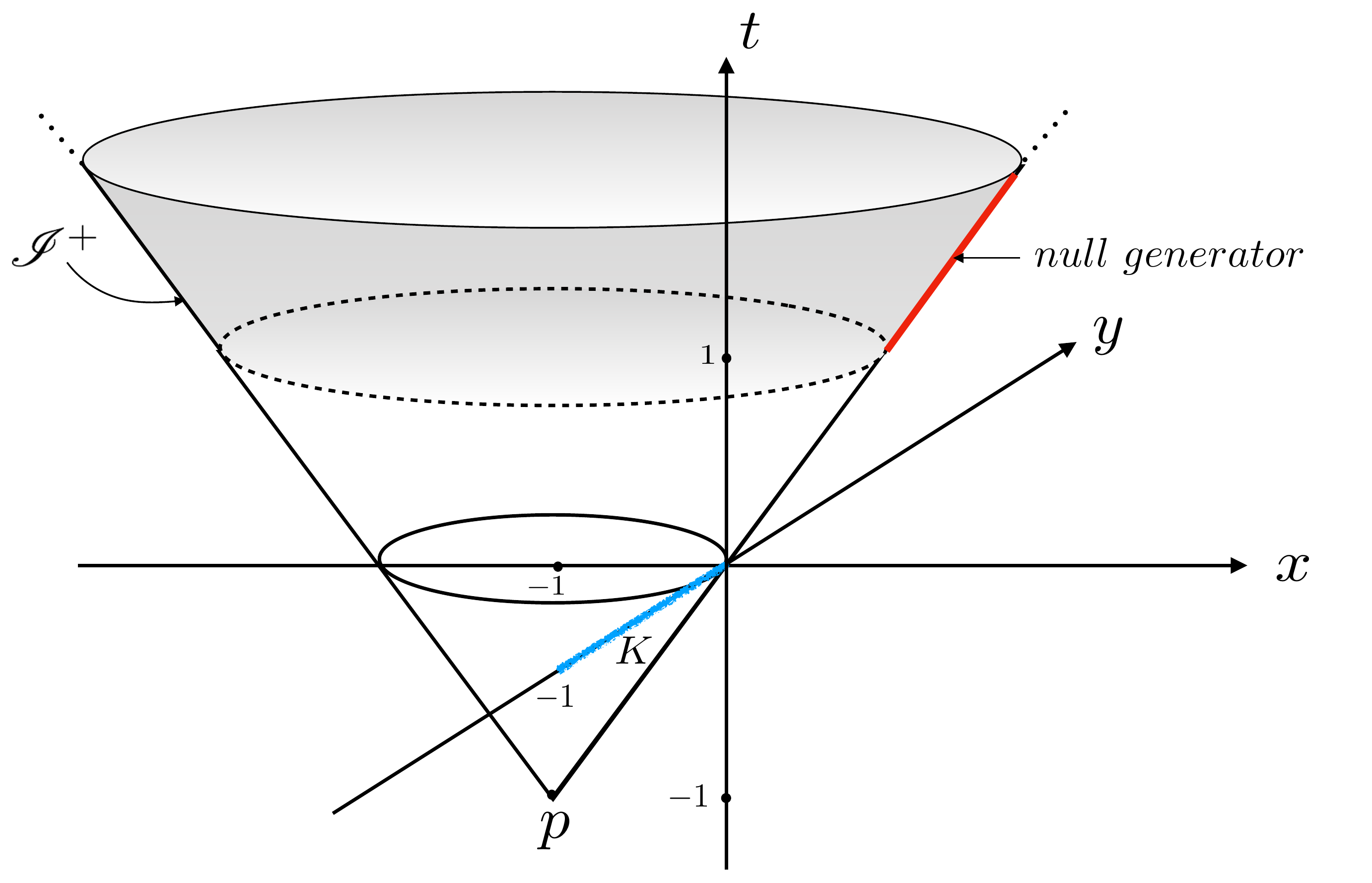}
}
\end{center}
\caption{\small $(\mcM,g)$ is causally simple but $(\bmcM,\tilde{g})$ is not.}
\label{example}
\end{figure}

\begin{example}
  \label{Ex5II17.2}
\rm
  It is easy to give an example  of a two-dimensional space-time  with a null $\scri^+$ which is causally simple but the conformal completion is not. On the other hand, it is rather clear that there are no  two-dimensional examples with ``causally simple''  
  replaced by ``future causally simple'' in the last sentence. But a three dimensional example  can be given, as follows: 
   Let  $p$ be the point in three dimensional Minkowski space-time $ \R^{1,2}$ with coordinates $(t,x,y)=(-1,-1,0)$,
  Let $(\mcM,g)\subset \R^{1,2}$ be obtained by removing the set $J^+(p)\cap \{t\ge 1\}$ from $\R^{1,2}$. Then $(\mcM,g)$ is causally simple. Let $(\bmcM,\tilde{g})$ be obtained by adding to $\mcM$ the null hypersurface 
  $$
   \scri^+:= \{t>1, (x+1)^2+y^2=(t+1)^2\}\equiv \{t>1\}\cap \dot J (p,\R^{1,2})  
   \,,
  $$
  with $\tilde g$ obtained by restriction of $g$; thus $(\bmcM,\tilde{g})$ has a boundary
  $
   \scri^+:=\partial \bmcM
   $
    which is that part of the Minkowskian light-cone of $p$ which lies to the timelike future of $\{t=1\}$. (One can multiply the flat metric on $\bmcM$ by a suitable conformal factor to comply with the usual definition of conformal completion af infinity, but this is clearly irrelevant for the problem at hand.)
Let 
$$
 K=\{t=0,x=0,y\in [-1,0] \}
  \,.
$$
Then $K$ is compact, and the null generator 
$$
 \{t>1\,\ x= t\,,\ y=0\}
$$
of $\scri^+$ is in the closure of $J^+(K,\bmcM)$ but is not in $J^+(K,\bmcM)$; cf.~Figure~\ref{example}.
Thus  $(\mcM,g)$ is future causally simple but  $(\bmcM,\tilde{g})$ is not.

\end{example}

\begin{example}
  \label{Ex5II17.1}
  {\rm
  The following example shows that Lemma~\ref{lemma} is wrong when $\scri^+$ is timelike:
  Let $(\mcM,g)\subset \R^{1,2}$ be the three-dimensional Minkowski space-time from which the quadrant $\{t\ge 1, x\le 0\}\times \R$ has been removed. Then $(\mcM,g)$ is again causally simple. Let $(\bmcM,\tilde{g})$ be obtained by adding to $\mcM$ a timelike future boundary $\scri^+:=\{t>1, x=0\}\times \R$.
(We ignore again the conformal factor.)  Let $p=(0,0,0)$. The null lines $\{t>1,x=0, y= \pm t\}$ are in the closure of $J^+(p,\bmcM)$ but are not in $J^+(p,\bmcM)$.
  }
\end{example}

\proof[Proof of Lemma \ref{lemma}]
Let $K \subset \mcM$ be a compact set. Let $q$ be a limit point  of $J^+(K, \bmcM)$. Then there is a sequence of points $\{q_n\} \subset J^+(K, \bmcM)$ such that $q_n \to q$, and so there is a sequence of future causal curves $\g_n \colon [0,1] \to \bmcM$ with $\g_n(0) = p_n$ and $\g_n(1) = q_n$. By restricting to a subsequence, we can assume that $p_n \to p \in K$. Using a normal neighborhood  centered at $q$ in $\scri^+$, we construct a compact spacelike hypersurface (with boundary) $\S \subset \mcM$ by pushing this neighborhood a small amount to the past along the past directed normal geodesics to $\scri^+$. Take $\S$ large enough so that, by passing to another subsequence if necessary, there exist unique points $r_n$ where $\g_n$ and $\S$ intersect. Since $\S$ is compact, there is a point $r \in \S$ such that $r$ is a limit point of $r_n$. So, since $(\mcM,g)$ is causally simple, there is a future causal curve $\l$ from $p$ to $r$. By choosing $\S$ sufficiently close to $\scri^+$, we may assume $r$ is in $D(\scri^+)$,  the domain of dependence of $\scri^+$. Then by applying \cite[Proposition 2.8.1]{Chru} in
$D(\scri^+)$,
there is a future causal curve $\s$ from $r$ to $q$. Concatenating $\l$ and $\s$ yields a future causal curve from $p$ to $q$.
\qed

\medskip 

We can form analogous statements for trapped regions with assumptions only on the outward expansion $\theta^+$. To be precise let $(\mcM,g)$ be a spacetime, and let $T$ be a compact connected spacelike hypersurface with boundary $S$ in $\mcM$.  If $S$ has null expansion $\theta^+\le 0$ with respect to the {\it outward} null normal vector field $\ell^+$ along $S$, then we say that  $T$ is a \emph{future weakly trapped region} and $S$ is a \emph{future weakly outer trapped surface}. Analogous statements for future  weakly outer trapped surfaces hold for Theorems \ref{main} and \ref{notrapped in dS}. For brevity we only state and prove the analogous version of Theorem \ref{notrapped in dS}.

\medskip

\begin{thm}
Consider a future asymptotically de Sitter spacetime  $(\mcM,g)$ which is future causally simple and  satisfies  the null energy condition.  Let $A \subset \mcM$ be such that $J^+(A, \bmcM)$ does not contain all of $\scri^+$.   Then there are no future weakly trapped regions contained in $J^+(A, \bmcM) \cap I^-(\scri^+, \bmcM)$.
\end{thm}

\proof
Suppose there is a future trapped region $T$, with boundary $S$, contained in $J^+(A, \bmcM) \cap I^-(\scri^+, \bmcM)$. As in the proof of Theorem \ref{main}, construct a null geodesic $\g\colon [a,b] \to \bmcM$ whose image is contained in
\[
\pd J^+(T, \bmcM) = J^+(T, \bmcM) \setminus I^+(T, \bmcM)  \,,
\]
 with $\g(a) \in T$ and $\g(b) \in \pd J^+(T, \bmcM) \cap \scri^+$. Since $\g$ does not enter the timelike future of $T$ we have $\g(a) \in S$ and $\g'(a)$ points in the outward direction. Let $\theta^+(s)$ be the null expansion along $\g(s)$ of the null hypersurface constructed as in the proof of Theorem \ref{main}.  Since $S$ is future weakly outer trapped and $(\mcM,g)$ satisfies the null energy condition, we have
 $\theta^+(s) \leq 0$ for all $s$. The rest of the proof is the same.
\qed

\section{Examples}\label{Examples}

\medskip

\noindent{\it De Sitter Space.}  Consider de Sitter space,
$$
\mcM = (-\pi/2, \pi/2) \times S^n \,, \quad g = \cos^{-2}(t)(-dt^2 + d\omega^2) \, ,
$$
where $d\omega^2$ is the usual round metric on the sphere $S^n$.
It conformally embeds into the Einstein static universe $(\mcM', g') = (\R \times S^n, -dt^2 + d\omega^2)$. The future conformal completion is $\bmcM = (-\pi/2, \pi/2] \times S^n$ with $\scri^+ = \{\pi/2\} \times S^n$.   Note that for any $p \in \mcM$, $J^+(p,\bmcM)$ does not cover all of $\scri^+$.  Hence, by Theorem \ref{notrapped in dS}
there are no future weakly trapped surfaces in $J^+(p,\mcM)$.

\begin{figure}[t]
\captionsetup{width=.85\linewidth}
\begin{center}
\mbox{
\includegraphics[width=4.3in]{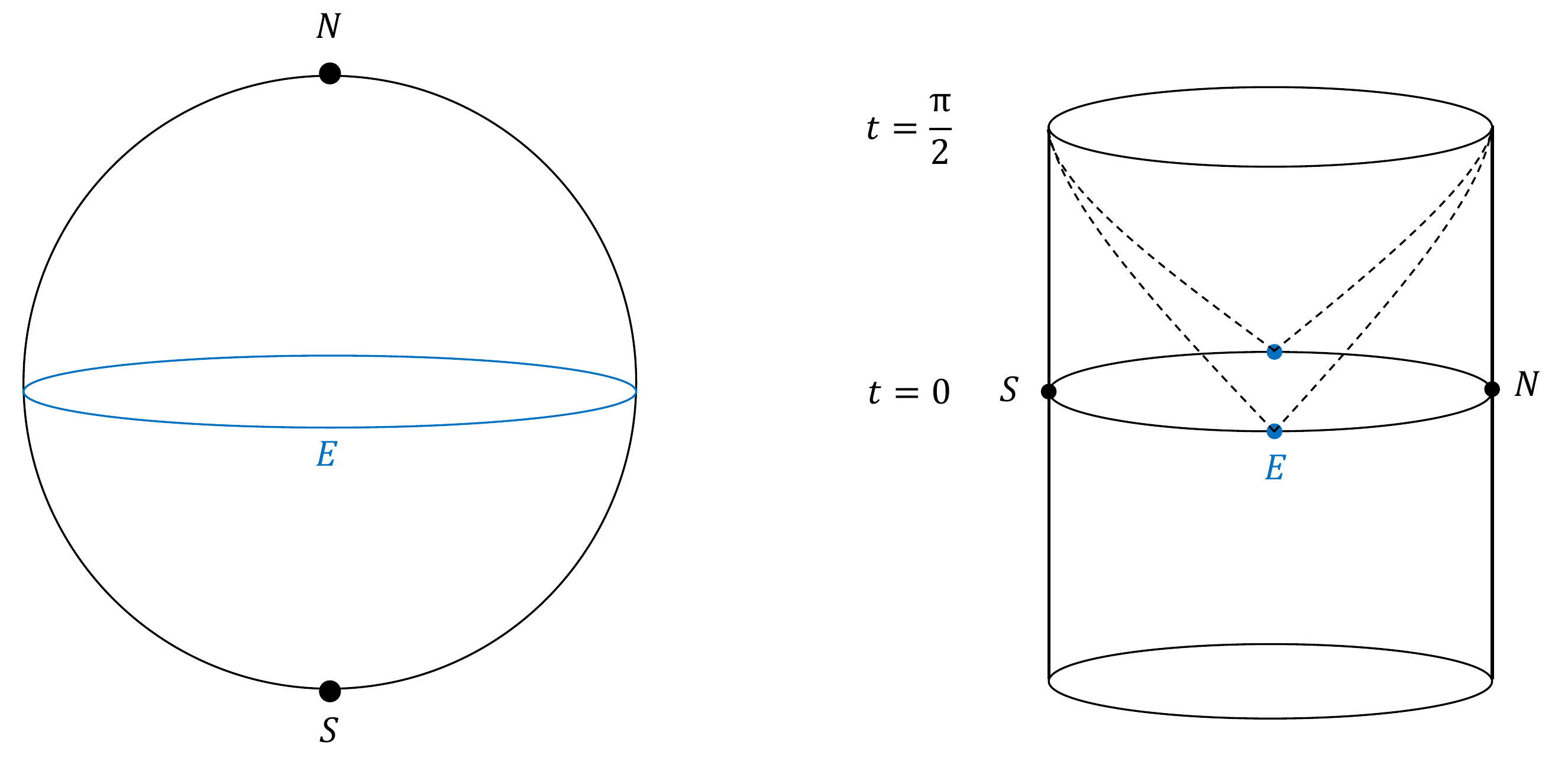}
}
\end{center}
\caption{\small{Left: An equator in the sphere $S^n$. Right: A spacetime diagram of de-Sitter space. Here the equator is represented by the two blue points. The causal future of $E$ covers all of $\scri^+ = \{t = \pi/2\}$, but the timelike future of $E$ misses the north and south pole points on $\scri^+$.}}
\label{dsfig1}
\end{figure}

Now consider any $t$-slice $\S_t = \{t\}\times S^n$. Let $K$ be the second fundamental form of $\S_t$ within $\mcM$. Let $S$ be any hypersurface in $\S_t$ and $H$ the mean curvature of $S$ within $\S_t$. Then
\beq\label{theta}
\theta^{\pm} = {\rm tr}_S K \pm H  \,,
\eeq
(With the notation as in the beginning of Section \ref{results}, here $\ell_{\pm}$ are scaled with respect to the unit normal $\nu$ tangent to the time slice.)  If $0 < t < \pi/2$, then $\text{tr}_{S}K > 0$ and so either $\theta^+$ or $\theta^-$ is positive. Consequently there are no future weakly trapped surfaces in
$\S_t$ for $t >0$.\footnote{In fact,
it can be shown that there are no future weakly trapped surfaces in the spacetime region $t >0$.} On the other hand, there are many future weakly trapped surfaces in $\S_0$. For example take $S = E$ to be the equator. Since $\S_0$ is totally geodesic within $\mcM$ and $E$ is a minimal surface within $\S_0$, we have $K = H = 0$. Thus, by \eqref{theta}, $\theta^{\pm} = 0$, and so $E$ is future weakly (in fact, marginally) trapped. This is consistent with Theorem \ref{notrapped in dS} since $J^+(E,\bmcM)$ contains all of $\scri^+$; see Figure~\ref{dsfig1}.  Let's also recognize that Theorem \ref{notrapped in dS} cannot be weakened by replacing `causal future' with `timelike future' because $I^+(E,\bmcM)$ does not contain all of $\scri^+$, since it misses the north pole and south pole associated with the equator.

\begin{figure}
\captionsetup{width=.85\linewidth}
\begin{center}
\mbox{
\includegraphics[width=4.7in]{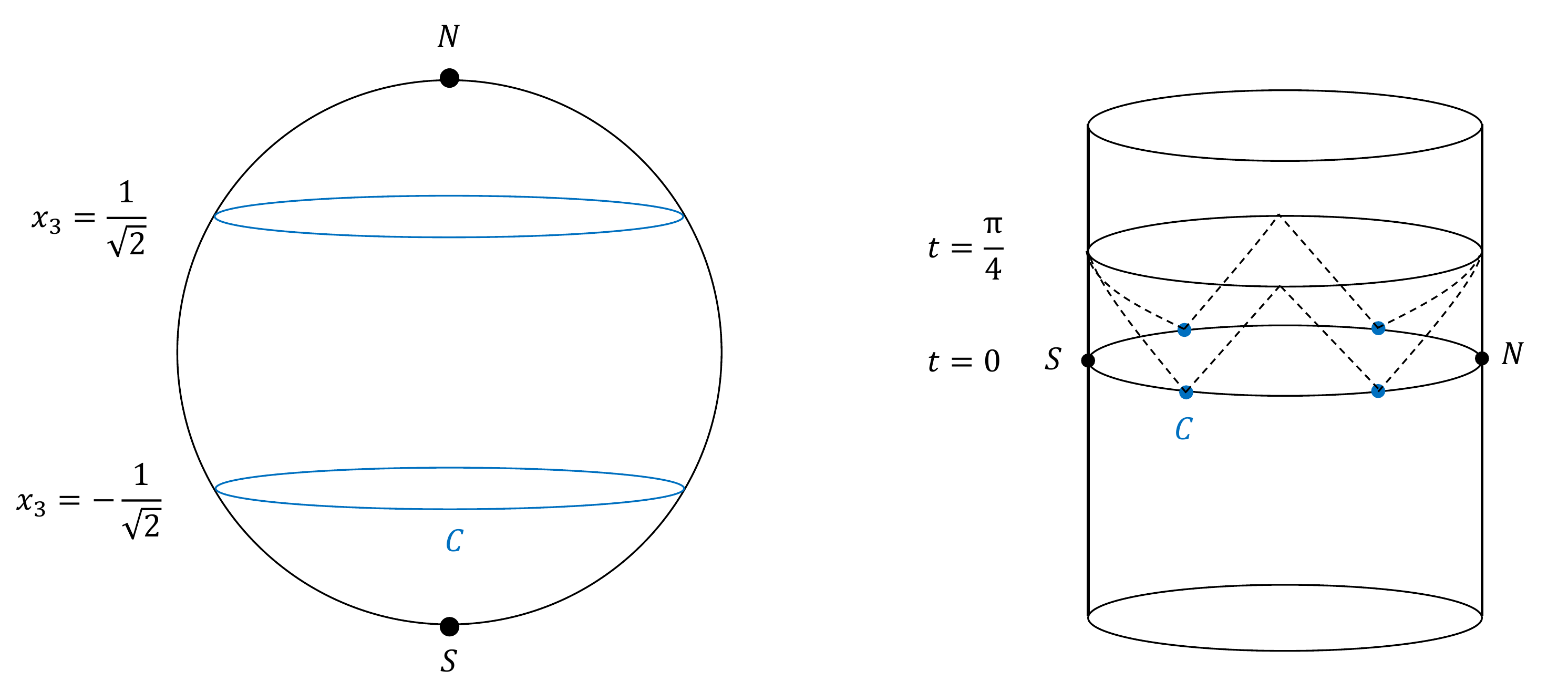}
}
\end{center}
\caption[2in]{\small{Left: A slice of the Clifford torus in the $x_4 = 0$ plane. Right: A spacetime diagram of de-Sitter space with $x_2 = x_4 = 0$. The Clifford torus is represented by the four blue points located at $(x_1, x_3) = (\pm 1/\sqrt{2}, \pm 1/\sqrt{2})$. The causal future of $C$ already covers the time slice $\{\frac{\pi}4\} \times S^3$.
}}
\label{dsfig2}
\end{figure}

The above example shows that {\it any} compact embedded minimal surface in $\S_0$ is a future weakly trapped surface (see \cite{Flores} for some related discussion).  But there are infinitely many such examples in the $3$-sphere, of arbitrary genus; see e.g. \cite{Brendle}.  (The maximum principle implies that any such example cannot be contained in an open hemisphere;  somewhat amusingly, this also follows from Theorem \ref{notrapped in dS}.) An interesting example in this case, where dim~$\mcM =4$, is the Clifford torus $C \subset \S_0$.  Expressing $S^3$ as the sphere in $\bbR^4$,
$$
x_1^2 + x_2^2 + x_3^2 + x_4^2 = 1 \,,
$$
the Clifford torus  is defined by the equations $x_1^2 + x_2^2 = 1/2 = x_3^2 + x_4^2$. The Clifford torus is a minimal surface, so one again has $\theta^{\pm} = 0$, and $C$ is a weakly trapped surface in
$\mcM$. This is consistent with Theorem \ref{notrapped in dS} since the causal future of $C$ covers all of $\scri^+$. In fact, the causal future of $C$ already   covers the time slice $\{\frac{\pi}4\} \times S^3$; see
Figure~\ref{dsfig2}.

\medskip

\begin{figure}[h]
\begin{center}
\mbox{
\includegraphics[width=5.3in]{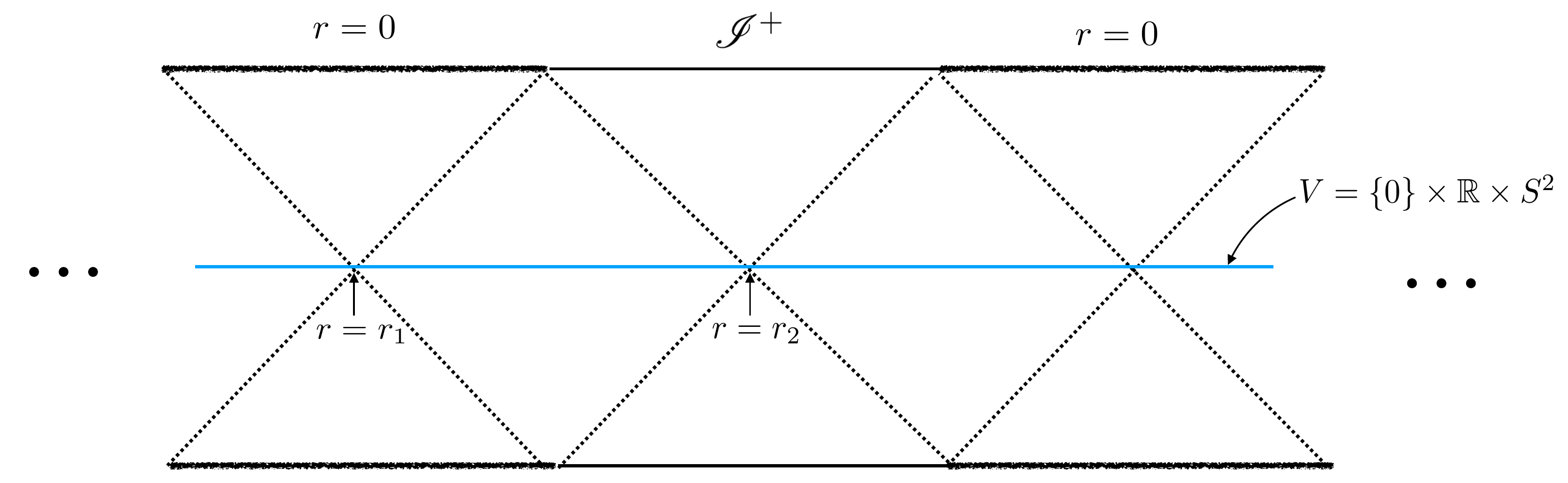}
}
\end{center}
\caption{Schwarzschild-de Sitter Space.}
\label{adsfig}
\end{figure}

\noindent
{\it Schwarzschild-de Sitter space.}  Schwarzschild-de Sitter space is the spacetime $\mcM = \R \times \R \times S^{2}$ with metric  in static form,
\beq\label{sdsg}
g = -\left(1 - \frac{2m}{r} - \frac{\Lambda }{3}r^2 \right)dt^2 + \left(1 - \frac{2m}{r} - \frac{\Lambda}{3}r^2 \right)^{-1}dr^2 + d\omega^2  \,,
\eeq
with $m > 0$, $\Lambda > 0$, and $9 \Lambda m^2 < 1$. The Penrose diagram for $(\mcM, g)$ is given in Figure~\ref{adsfig}; cf.\  \cite{GibHawk}.
$\scri^+$ has topology $\R \times S^{2}$.  (Here we consider a conformal compactification consisting of a single component of future null infinity.)  The $g_{tt}$-component of  \eqref{sdsg} has positive roots $r_1 < r_2$, corresponding to a black hole event horizon and a cosmological horizon, respectively.  Regular Kruskal-Szekeres type coordinates can be defined near $r = r_1$ and $r = r_2$.

Consider the totally geodesic time slice $V =\{0\} \times \R \times S^{2}$.   Let $A$ be the subset of  $V$
consisting of the union of all rotationally symmetric $2$-spheres $S_r$, for $r_1 <  r \le r_0$.
Provided $r_0 < r_2$,  Theorem \ref{notrapped in dS} implies that there are no future weakly  trapped surfaces in $J^+(A, \bmcM) \cap I^-(\scri^+, \bmcM)$.  In particular, we note that the $2$-spheres $S_r$  have positive null expansion with respect to the null normal pointing to the right.  However, when we allow $r_0 = r_2$, then $J^+(A, \bmcM) \cap I^-(\scri^+, \bmcM)$ contains the $2$-sphere at $r = r_2$, which is future weakly trapped, since it is minimal.  Again this is consistent with Theorem \ref{notrapped in dS}, since now $J^+(A)$ contains all of $\scri^+$.

\medskip

\noindent
\textsc{Acknowledgements}  The research of PTC  was  supported
by the Polish National Center of Science (NCN) under grant 2016/21/B/ST1/00940.  The research of GJG was supported by the NSF under the grant DMS-171080.



\end{document}